\documentclass[prc,twocolumn,letterpaper,10pt,twoside,tightenlines,nofootinbib,showpacs]{revtex4-1}
\usepackage{amsmath,amsfonts,amssymb,latexsym}
\usepackage{graphicx}
\usepackage[sort&compress]{natbib}
\usepackage{grffile,epstopdf}

\newcommand{\m}{\cdot}

\newcommand{\n}{\nonumber \\}

\newcommand{\llkl}{\left\langle}
\newcommand{\rrkl}{\right\rangle}

\newcommand{\kl}{\left(}
\newcommand{\kr}{\right)}

\newcommand{\td}{\text{d}}

\newcommand{\lv}{\left\vert}
\newcommand{\rv}{\right\vert}
\newcommand{\sel}{\sigma_\text{el}}
\newcommand{\nab}{\vec{\bigtriangledown}}
\newcommand{\del}{\partial}

\makeindex

\begin{document}
\title{Electric Conductivity of the Quark-Gluon Plasma investigated using a perturbative QCD based parton cascade}

\author{Moritz Greif}
\email{greif@th.physik.uni-frankfurt.de}
\affiliation{Institut f\"ur Theoretische Physik,
Johann Wolfgang Goethe-Universit\"at,
Max-von-Laue-Str.\ 1, D-60438 Frankfurt am Main, Germany}

\author{Ioannis Bouras}
\affiliation{Institut f\"ur Theoretische Physik,
Johann Wolfgang Goethe-Universit\"at,
Max-von-Laue-Str.\ 1, D-60438 Frankfurt am Main, Germany}

\author{Zhe Xu}
\affiliation{Department of Physics, Tsinghua University, Beijing 100084, China and Collaborative Innovation Center of Quantum Matter, Beijing 100084, China}

\author{Carsten Greiner}
\affiliation{Institut f\"ur Theoretische Physik,
Johann Wolfgang Goethe-Universit\"at,
Max-von-Laue-Str.\ 1, D-60438 Frankfurt am Main, Germany}

\date{\today}


\begin{abstract}
Electric conductivity is sensitive to effective cross sections among the particles of the partonic medium. We investigate the electric conductivity of a hot plasma of quarks and gluons,  solving the  relativistic Boltzmann equation. In order to extract this transport coefficient, we employ the Green-Kubo formalism and, independently, a method motivated by the classical definition of electric conductivity. To this end we evaluate the static electric diffusion current upon the influence of an electric field. Both methods give identical results. For the first time, we obtain numerically the Drude electric conductivity formula for an ultrarelativistic gas of quarks and gluons employing constant isotropic binary cross sections.
 Furthermore, we extract the electric conductivity for a system of massless quarks and gluons including screened binary and inelastic, radiative $2\leftrightarrow 3$ perturbative QCD scattering. Comparing with recent lattice  results, we find an agreement in the temperature dependence of the conductivity.
\end{abstract}

\maketitle

\section{Introduction}
\label{sec:Intro}
Ultrarelativistic collisions of heavy nuclei generate a temporary state of matter, called quark-gluon plasma (QGP)~\cite{Gyulassy2005}, in which quarks and gluons are the relevant degrees of freedom. Experimentalists measure sensible observables with high precision in order to characterize the key properties of nature on the smallest scale. Theoretical frameworks are necessary to gain physical insight when compared to experimental data. Many steps are required to achieve this goal; therefore, it is of importance to investigate theoretical models and theories, even if the outcome is not directly comparable with data. 
Among the best examples are relativistic transport coefficients, which cannot be measured directly. 
Nevertheless, it is important to employ hydrodynamic models, where viscous corrections are taken into account ~\cite{GALE2013,Schenke2011}. Especially the existence of a finite shear viscosity in the QGP is necessary to explain experimental data of the elliptic flow coefficient $v_2$ ~\cite{Schenke2011a}. 
In some cases, transport coefficients can be calculated analytically ~\cite{Denicol2012b,Denicol2012,Denicol2010}, for example, using constant isotropic cross sections. 
For more realistic scenarios, the extraction of transport coefficients requires numerically solvable theories, such as the relativistic Boltzmann transport theory.
In the past, the numerical solution of the Boltzmann equation was successfully applied to extract the shear viscosity over entropy ratio $\eta/s$ ~\cite{Wesp2011, Reining2012,Plumari2012}, as well as the heat conductivity coefficient $\kappa$ numerically ~\cite{Greif2013}.
Heat flow, shear viscosity and bulk viscosity are coefficients that also appear naturally in kinetic theory of single-component systems, even without external forces.
The electric conductivity is only well defined in systems where at least two differently charged particle types are present and can scatter with each other.
The longitudinal static electric conductivity $\sel$ relates the response of the electrically charged particle diffusion current density $\vec{j}$ to an externally applied static electric field $\vec{E}$:
\begin{equation}
\vec{j}=\sigma_{\text{el}}\vec{E}.
\end{equation}
Recently, several scientific groups focused on this transport coefficient ~\cite{Arnold2000,Arnold2003,Gupta2004,Aarts2007,Buividovich2010,Ding2011,Burnier2012,Brandt2013,Amato2013a,Cassing2013,Steinert2013,Puglisi2014a,Finazzo2014}.
The electric conductivity is related to the soft dilepton production rate ~\cite{Moore2003} and the diffusion of magnetic fields in the medium ~\cite{Baym1997,Tuchin2013,Fernandez-Fraile2006}. Indeed, it provides us with the possibility to compare effective cross sections of a medium's constituents among several theories, including transport models ~\cite{Cassing2013,Steinert2013}, lattice gauge theory ~\cite{Aarts2007,Brandt2013,Amato2013a,Gupta2004,Buividovich2010,Burnier2012,Ding2011} and Dyson-Schwinger calculations ~\cite{Qin2013}. 

In this work we present a systematic study of the electric conductivity coefficient for a hot plasma governed by perturbative quantum chromodynamics (pQCD), applying the microscopic relativistic transport model \textit{Boltzmann Approach to Multi-Parton Scatterings} (BAMPS) ~\cite{Xu2005,Bouras2010a,Bouras2012,Bouras2009,Fochler2010,Fochler2011,Uphoff2011a,Wesp2011,Reining2012,Uphoff2012,Fochler2013,Greif2013,Senzel2013,Uphoff2013}. 
We set out to investigate the effects of elastic and inelastic scattering, with or without running coupling. This allows us to gain insights in effective scattering rates from lattice QCD (lQCD) by comparing results for the electric conductivity with those from the transport calculation BAMPS.
This work is organized as follows.
In Sec.~\ref{basic_definitions} we give basic definitions regarding the relativistic formulation for the fluid dynamical quantities. In Sec.~\ref{sec:BAMPS} we present the framework for the numerical solution of the Boltzmann equation, BAMPS. In order to obtain the electric conductivity we use two different methods, the Green-Kubo formula using correlation functions (Sec.~\ref{sec:Extraction-of-the-tr-coeff}) and small electric fields (Sec.~\ref{sec:electric-field}). We show that both methods coincide and in Sec.~\ref{sec:NumericalResultsfixedCS} we compare results from BAMPS to analytic formulas. In Sec.~\ref{sec:pqcd_systems_results} we present our results for full inelastic pQCD cross sections, and contrast them with previous investigations from other groups. We give a conclusion and outlook in Sec.~\ref{conclusion}. 
Our units are $\hbar =c=k=1$; the space-time metric is given by $g^{\mu \nu
}=\text{diag}(1,-1,-1,-1)$. Greek indices run from $0$ to $3$.
\section{Basic definitions}
\label{basic_definitions}
In relativistic fluid dynamics a system is described by the energy-momentum tensor and the four-currents of conserved charges. We consider a system of $k=\lbrace 1,\ldots,M\rbrace$ particle species, each of which carries the electric charge $q_k$. The phase-space distribution function is named $f_k(x,p)$.
Defining the components of the \textit{particle} flow of species $k$ by 
\begin{equation}
N_k^\mu(x)=\int \frac{g_k\,d^3p_k}{(2\pi)^3p_k^0}\, p_k^\mu f_k(x,p_k),
\end{equation}
the total particle flow is
\begin{equation}
N^\mu = \sum\limits_{k=1}^M N_k^\mu.
\end{equation}
Here $p_k^\mu=(E_k,\vec{p}_k)$ is the four-momentum of the particle of species $k$ and $g_k$ is its degeneracy factor.
We introduce the four-velocity as an arbitrary normalized timelike four-vector $u^\mu=\gamma(1,\vec{v})$, where $u_\mu u^\mu=1$.
The particle flow is decomposed into a part orthogonal to the four-velocity, the \textit{diffusion} flow of species $k$, $V^\mu_k$, and one part parallel to it,
\begin{equation}
N_k^\mu(x)=n_k u^\mu + V^\mu_k.\label{Nmuformel}
\end{equation}
Here we define the local rest frame (LRF) number density of species $k$ as 
\begin{equation}
n_k = N^\mu_k u_\mu.
\end{equation}
The total LRF particle number density is 
\begin{equation}
n\equiv \sum\limits_{k=1}^M n_k,
\end{equation}
and the density fraction of species $k$ compared to the total density is defined as
\begin{equation}
x_k\equiv \frac{n_k}{n}.
\end{equation}
Using the spatial projector $\Delta^{\mu\nu}=g^{\mu\nu}-u^\mu u^\nu$, the diffusion flow of species $k$ reads
\begin{equation}
V^\mu_k(x)=\Delta^\mu_\nu N^\nu_k(x).
\end{equation}
Without loss of generality, we can use the Eckart definition of the four-velocity, $u^\mu=N^\mu/\sqrt{N_\mu N^\mu}$. Then another useful form of the particle diffusion current can be written as
\begin{equation}
V_k^\mu = N^\mu_k - x_k N^\mu.
\label{DiffusionFlow}
\end{equation}
From Eq.~\eqref{DiffusionFlow}, it is clear that the diffusion flow of a particle species is the particle flow of this species with respect to the scaled total flow of the system.
The total electric current density is defined as
\begin{equation}
j^\mu=\sum\limits_{k=1}^M q_k V^\mu_k.
\end{equation}
Note that in the LRF, Eq.~\eqref{Nmuformel} simplifies to $V^i_k = N^i_k$ for spatial components $i={1,2,3}$.
\section{The partonic cascade BAMPS}
\label{sec:BAMPS}
In this work, the relativistic 3+1-dimensional Boltzmann equation is solved numerically using the semiclassical parton cascade BAMPS, developed
and previously employed in Refs.~~\cite{Xu2005,Bouras2009,Fochler2010,Fochler2011,Uphoff2011a,Wesp2011,Reining2012,Uphoff2012,Fochler2013,Greif2013,Senzel2013,Uphoff2013,Bouras2010a}.
BAMPS solves the Boltzmann equation microscopically, 
\begin{equation}
\left(\frac{\del}{\del t} + \frac{\vec{p}}{E}\m\nab \right)f_k(x,t)=\mathcal{C}_k^{\rm 2 \rightarrow 2}\left[ f\right]+\mathcal{C}_k^{\rm 2 \leftrightarrow 3}\left[ f\right]+ \cdots,
\label{eq:BTE}
\end{equation}
for on-shell particles using the stochastic interpretation of transition
rates. The left-hand side of Eq.~\eqref{eq:BTE} describes the evolution of the single-particle distribution function $f_k$ of species $k$. All $\rm 2\rightarrow 2$ and $\rm 2\leftrightarrow 3$ processes for massless light quarks ($q$) and gluons ($g$) are included. The right-hand side describes the interactions (collisions) between the particles: $\mathcal{C}_k^{\rm 2 \rightarrow 2}$ refers to the elastic collision term of the partons, whereas $\mathcal{C}_k^{\rm 2 \leftrightarrow 3}$ describes inelastic processes. 
In order to reduce statistical fluctuations in simulations and to ensure an accurate solution of the Boltzmann equation \eqref{eq:BTE}
the widely used test particle method ~\cite{Xu2005} is introduced: The particle number $N$ 
is artificially increased by multiplying it by the number of
test particles per real particle, $N_{\rm test}$:
\begin{equation}
N\rightarrow N_{\rm test}\m N
\end{equation}
 The physical results
are not affected by this procedure, because the cross sections $\sigma$ are scaled simultaneously,
\begin{equation}
\sigma\rightarrow \sigma/N_{\rm test}.
\end{equation}
Throughout this work, we use BAMPS as a multiparticle simulation for up, down and strange quarks, their corresponding antiquarks and gluons. All particles have physical degeneracies and charges, they are on shell and massless.
Debye masses are dynamically computed within BAMPS along the current parton distribution and solely applied to cure IR divergences occurring in the integrations of the matrix elements ~\cite{Xu2005}. All on-shell particles remain massless. 
All BAMPS setups are electrically neutral, because we always initialize as many antiparticles as particles.
Space in BAMPS is discretized into suffiently small volume elements (cells) with volume $V_{c}$, while time is discretized in steps $\Delta t$. The cells are populated by particles, that scatter with each other stochastically, depending on the scattering rate. The rates are computed in leading-order pQCD, or alternatively, using a fixed cross section. 

In this paper, the solution of the relativistic Boltzmann equation is obtained for a quadratic, static box of volume $V=L_xL_yL_z$. We employ periodic boundary conditions.

The distribution function $f(x,p)$ of each volume element is reconstructed from the momenta distribution of the particles inside it. In this scheme, the particle flow $N^{\mu }$ and energy-momentum tensor $T^{\mu \nu }$ are computed via the discrete
summation over all particles within the specific volume element and divided
by the test particle number,
\begin{align}
N^{\mu }(t,x)& =\frac{1}{V_{c} N_{\rm test}}\sum\limits_{i=1}^{N_{c}}\frac{p_{i}^{\mu }}{%
p_{i}^{0}},  \label{eq:nmu_BAMPS} \\
T^{\mu \nu }(t,x)& =\frac{1}{V_{c} N_{\rm test}}\sum\limits_{i=1}^{N_{c}}\frac{p_{i}^{\mu
}p_{i}^{\nu }}{p_{i}^{0}},  \label{eq:tmunu_BAMPS}
\end{align}%
where $N_{c}$ is the total number of particles inside the corresponding
cell, $t$ is the time, and $x$ is the space coordinate (defined to be located in
the center of the volume element).
In this study, we first consider only binary collisions with constant isotropic
cross sections (Secs. \ref{sec:NumericalMethods} and \ref{sec:NumericalResultsfixedCS}). 

Afterwards, we employ elastic and also inelastic pQCD cross sections (Sec.~\ref{sec:pqcd_systems_results}). With total cross sections $\sigma_{22},\sigma_{23}$ for $2\rightarrow 2, 2\rightarrow 3$ collisions, the collision probabilities for two particles inside a grid cell of volume $V_{c}$ within a time step $\Delta t$ are
\begin{equation}
P_{22,23}=v_\text{rel}\frac{\sigma_{22,23}}{N_\text{test}}\frac{\Delta t}{V_c}
\label{eq:Probab22}
\end{equation}
and accordingly for the inelastic backreaction $3\rightarrow 2$,
\begin{equation}
P_{32}=\frac{1}{8E_1E_2E_3}\frac{I_{32}}{N_\text{test}^2}\frac{\Delta t}{V^2_{c}}
\label{eq:Probab32}
\end{equation}
where $I_{32}$ is given by an integral over the final states of the interaction process, and corresponds to a cross section for $3\rightarrow 2$ scattering, and $v_\text{rel}=(p_1+p_2)^2/(2E_1E_2)$ is the relative velocity of the two incoming massless particles with four-momenta $p_{1,2}=(E_{1,2},\vec{p}_{1,2})$.

The matrix elements underlying the elastic collision cross sections are calculated in pQCD leading-order. The inelastic cross sections are obtained through the Gunion-Bertsch matrix element ~\cite{Gunion1982}, which was further improved ~\cite{Fochler2013} and applied within BAMPS computations in Ref.~~\cite{Uphoff2014a}. The matrix element in the Gunion-Bertsch approximation factorizes into an elastic part $\mathcal{M}_{X\rightarrow Y}$ and a probability $P_g$ for the emission of a gluon,
\begin{equation}
\lv \mathcal{M}_{X\rightarrow Y+g}\rv^2=\lv \mathcal{M}_{X\rightarrow Y}\rv^2 P_g\label{eq:improved_gunion_bertsch}
\end{equation}
with
\begin{align}
P_g&=48\pi\alpha_s(k^2_\bot)(1-\bar{x})^2\n
&\ \times\left[\frac{\vec{k}_\bot}{k^2_\bot}+\frac{\vec{q}_\bot-\vec{k}_\bot}{\kl \vec{q}_\bot-\vec{k}_\bot\kr^2+m_D^2\kl \alpha_s(k^2_\bot)\kr}\right],
\end{align}
where $k_\bot$ and $q_\bot$ are the transverse momentum of the emitted and internal gluons, respectively. The longitudinal momentum fraction $\bar{x}$ carried away by the emitted gluon can be related to the gluon rapidity $y$ in the center-of-mass system of the respective microscopic collision by $\bar{x}=e^{\lv y\rv}k_\bot/\sqrt{s}$.
Within BAMPS, the running of the strong coupling $\alpha_s$ is evaluated explicitly   at the microscopic scale of the momentum transfer of the respective channel ~\cite{Uphoff2011a, Uphoff2012,Uphoff2014a}.
To account for the Landau-Pomeranchuk-Migdal (LPM) effect, describing the suppression of gluon emission due to finite gluon formation times $\tau_f$, an effective cutoff is implemented in the inelastic matrix elements, using the theta function ~\cite{Uphoff2014a}
\begin{equation}
\theta\kl \lambda - X_\text{LPM}\tau_f\kr.
\end{equation} 
The cutoff ensures that the mean free path $\lambda$ of the emitting parton exceeds the formation time of the emitted gluon times a phenomenological scaling factor, $X_\text{LPM}\tau_f$. Setting $X_\text{LPM}=0$ inhibits any LPM suppression whereas $X_\text{LPM}=1$ suppresses the gluon emission too strongly. 

Following Ref.~~\cite{Uphoff2014a}, the LPM cutoff parameter is fixed to $X_\text{LPM}=0.3$ in order to describe RHIC data of the nuclear modification factor $R_{AA}$ using BAMPS for full simulations of the partonic phase of heavy-ion collisions. Employing this parameter, the experimental data of the elliptic flow and nuclear modification factor at the LHC can also be understood on a microscopic level ~\cite{Uphoff2014a}.
\section{Numerical methods to extract the transport coefficient $\sel$}
\label{sec:NumericalMethods}
We present two independent methods to extract the electric conductivity $\sel$ from the partonic cascade BAMPS  and show the equivalence of the results. All results from this section are obtained using constant isotropic binary cross sections ($3~\mathrm{mb}$). 
\subsection{Green-Kubo relations}
\label{sec:Extraction-of-the-tr-coeff}
In an equilibrated system with volume $V$ and inverse temperature $\beta=T^{-1}$, the zero-frequency Green-Kubo ~\cite{Green1954,Kubo1957} formula 
for the electric conductivity is
\begin{equation}
\sel=\beta V \int\limits_0^\infty \llkl j_i(0)j_i(t) \rrkl \td t\label{WichtigeFormel},
\end{equation}
where $i=1,2,3$ represents the spatial direction, and no sum over $i$ is implied.
The electric current autocorrelation function $\llkl j_i(0)j_i(t) \rrkl$ can be obtained numerically, as done in Refs.~~\cite{Wesp2011, Demir2009} for the shear stress tensor correlation function. 
In general, for any time-dependent variable $A(t)$ known only for discrete, equally distributed  time steps $t=\left\lbrace t_0,t_1,\ldots,t_K\right\rbrace$, the autocorrelation function can only approximately be computed. If the value of the variable were available for infinitely many time steps, it could be fully computed. For a static system, the autocorrelation function is classically
\begin{equation}
C(t_l)=\frac{1}{s_{\text{max}}}\sum\limits_{s=0}^{s_\text{max}}A(t_s)A(t_s+t_l),\quad s_\text{max}=K-l.
\label{eq:classical_correlation_function}
\end{equation}
In the LRF of the fluid, the electric current density for systems of $M$ particle species with $N_k$ particles of species $k$ in the box reads
\begin{equation}
j^l(t)=\frac{1}{VN_{\text{test}}}\sum\limits_{k=1}^M q_k \sum\limits_{i=1}^{N_k} \left.\frac{p^l_i}{p_i^0}\right\vert_t.\label{eq:current_to_be_correlated}
\end{equation}
In our case we obtain $j^l(t)$ for all equivalent space directions $l=1,2,3$ from BAMPS.
Then it is possible to extract the autocorrelation function $C(t)$ of the electric current density in equilibrium.

\begin{figure}
\centering
\includegraphics[width=\columnwidth]{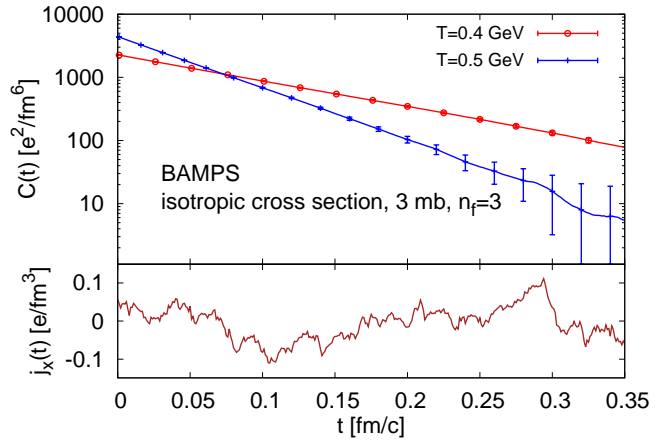}
\caption{Top: Examples of electric current autocorrelation functions $C(t)$ at different temperatures, with the standard error over the full sample (210 runs) shown for selected points. Bottom: Electric current fluctuation over time from a single run.}
\label{Correlator-Example}
\end{figure}

\begin{figure}
\centering
\includegraphics[width=\columnwidth]{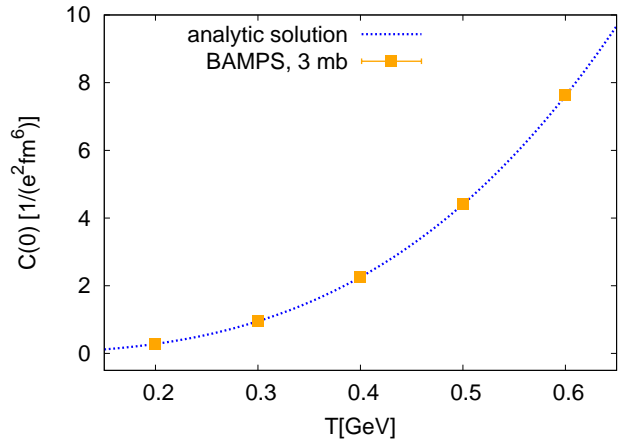}
\caption{The variance of the correlation function from the analytic formula from Eq.~\eqref{eq:variance_analytic_main_text} compared to numerical results, using constant isotropic binary cross sections.}
\label{VarianceVergleich}
\end{figure}

In order to get the value of $\sel$ from Eq.~\eqref{WichtigeFormel}, we integrate the numerically obtained electric current autocorrelation function.
For solutions of the Boltzmann equation, the autocorrelator has an exponential shape; see Refs.~\cite{Reichl2009,Wesp2011}.
We include this additional information in the analysis, as has been done previously for the case of shear viscosity~\cite{Wesp2011, Demir2009, Plumari2012}.
In summary, we fit an exponential function $C(t)=C(0)e^{-t/\tau}$ to the numerical correlator data and obtain the value of the variance $C(0)$ and the relaxation time $\tau$.
See Fig.~\ref{Correlator-Example} for a typical fluctuation of the equilibrium electric current and examples of the correlation function at two different temperatures. Clearly, the different slopes and variances $C(0)$ are visible. 
All the nontrivial information about the dynamics of the system is encoded in the relaxation time of the correlation function. For a system of volume $V$ containing $k=\left\lbrace 1,\ldots,M\right\rbrace$ particle species with charge $q_k$ and particle density $n_k$, the variance can even be computed analytically,
\begin{align}
C(0)=\frac{1}{3V}\sum\limits_{k=1}^M q_k^2 n_k
\label{eq:variance_analytic_main_text}
\end{align}
and is discussed in Appendix~\ref{appsec:Variancesection}.
We use the analytic value in all calculations, eliminating a source of numerical error. To justify this method, we show in Fig.~\ref{VarianceVergleich} the nice agreement of Eq.~\eqref{eq:variance_analytic_main_text} to several BAMPS results.

Using Eq.~\eqref{WichtigeFormel}, we obtain
\begin{equation}
\sel=\beta V C(0)\tau=\frac{\beta}{3}\left(\sum\limits_{k=1}^M q_k^2 n_k\right)\tau.
\label{eq:AnalyticVariancePluggedIn}
\end{equation}
We note that the statistical error of the fit parameter $\tau$ encounters the same difficulties as, e.g., in lQCD computations. A naive error estimation of the slope of the correlation function underestimates the error.
Here we have $N_\text{run}$ independent data sets from BAMPS; the runtime is $t_\text{run}$. The (equilibrium) setups are identical, but fluctuate independently. For each simulation, the $\llkl j_i(t)j_i(0) \rrkl $ correlator is calculated classically, along Eq.~\eqref{eq:classical_correlation_function}, for all spatial directions. The fluctuations in all three directions are completely independent; therefore, we have effectively $3N_\text{run}$ independent runs. 
From the sample of $3N_\text{run}$ correlators, we generate $3N_\text{run}$ new samples by always omitting one sample member of the original sample, and fit an exponential function $C(t)=C(0)e^{-t/\tau}$ to the average of each reduced sample. In this way we perform the well-known jackknife analysis ~\cite{Gattringer_Book} to obtain meaningful errors. The error decreases with decreasing ratio $\tau/t_\text{run}$ and increasing collision number per time step $\Delta t$. In Fig.~\ref{Correlator-Example}, the errors are standard errors of the full sample.
\subsection{Electric field}
\label{sec:electric-field}
\begin{figure}[t!]
  \centering
\includegraphics[width=\columnwidth]{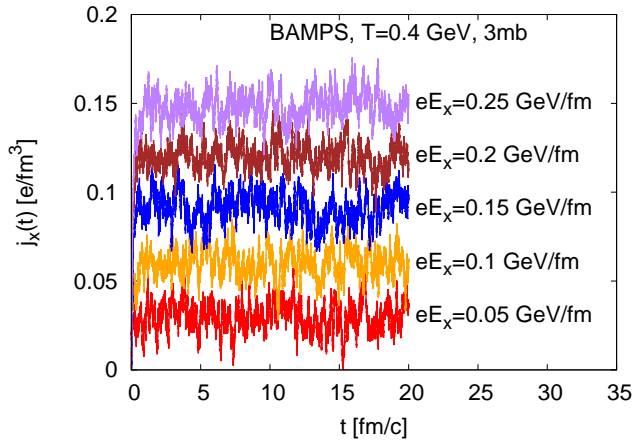}
\caption{The typical picture of static electric current densities, established by constant electric fields of different strengths. The BAMPS calculations use a constant isotropic cross section of $3~\mathrm{mb}$ and a temperature of $0.4~\mathrm{GeV}$. Here we show the average of 70 simulations.}\label{Electricfield1}
\end{figure}

\begin{figure}[t!]
  \centering
\includegraphics[width=\columnwidth]{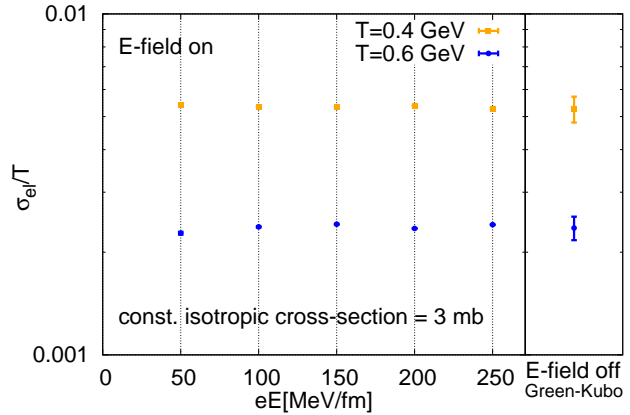}
\caption{For different electric field strengths, the electric conductivity is calculated and averaged. The standard error of the averaged currents is very small. For comparison, the corresponding Green-Kubo value is also plotted.}\label{Electricfield2}
\end{figure}

A straightforward method to extract the electric conductivity of the quark-gluon plasma is to run a box simulation in equilibrium and suddenly turn on a small and static electric field in $x$ direction, $\vec{E}=E\vec{e}_x$. Because of momentum transfer due to collisions a finite electric current will establish after a sufficiently large time. The average magnitude of this current is proportional to the electric field, and defines the zero-frequency electric conductivity $\sel$ directly,
\begin{equation}
\llkl j^x(t) \rrkl_{\text{static}} =\sel E^x.
\end{equation}
This method has been applied previously by the authors of Refs.~\cite{Steinert2013,Cassing2013}.
Numerically, the particle momenta for the individual particles in $x$ direction $p_i^x$  are influenced by the electric field in each numerical time step $\Delta t$,
\begin{equation}
p_i^x(t+\Delta t)\ \longrightarrow\ p_i^x(t)+ \Delta t E^x q_i, 
\end{equation}
with the individual electric charge $q_i$ of all the $i={1,\ldots,N}$ particles in the system. 
In Fig.~\ref{Electricfield1}, examples of static currents for different electric field strengths are shown. In Fig.~\ref{Electricfield2}, we show the electric conductivity for different electric field strengths and temperatures with the result of the Green-Kubo analysis. It can be seen that the results of both methods are compatible within the errors. It has been checked that the increase of temperature in the system is sufficiently small.

\section{Numerical results for fixed cross sections}
\label{sec:NumericalResultsfixedCS}
As discussed in the previous section, we are able to extract the electric conductivity reliably from the partonic cascade BAMPS with two independent methods. In the following we compare results of the electric conductivity with constant isotropic cross sections. In this case, we can compare with the relativistic generalization of the simple Drude formula ~\cite{Reif2009}. More refined analytic expressions for $\sel$, solving the linearized Boltzmann equation, remain as a future project.
\begin{figure}[t!]
  \centering
\includegraphics[width=\columnwidth]{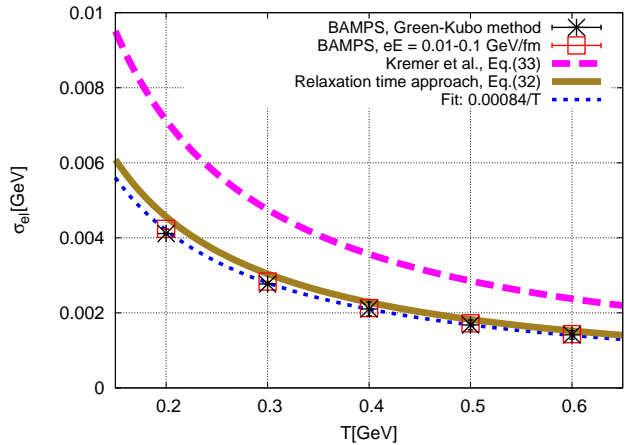}
\caption{Numerical results for the electric conductivity compared to analytic formulas. A constant isotropic binary cross section of $3~\mathrm{mb}$ is used in the system. Both methods from Sec.~\ref{sec:NumericalMethods} show identical results.}\label{BampsvsAnalytics}
\end{figure}
To compare with analytic formulas, it is useful to set the total cross section to a fixed value for all collisions ($3~\mathrm{mb}$), and let the (massless) particles scatter isotropically. 

Nonrelativistically, the Drude formula for the electric conductivity $\sigma_{\text{el,nr}}$ of a single charge-carrying species (e.g., electrons) with charge $q$, density $n_q$ and mass $m_q$ reads
\begin{equation}
\sigma_{\text{el,nr}}=\frac{n_q q^2 \tau}{m_q},
\label{Non-relativistic-Drude}
\end{equation}
where $\tau$ is the mean time between collisions of the charge carriers with, e.g., atomic cores.

Relativistically, the Boltzmann equation can be solved analytically in the relaxation time approximation, which is a simplistic model for the collision term. For this purpose, the Anderson-Witting model ~\cite{Anderson1974} for the collision term in Eq.~\eqref{eq:BTE} is used,
\begin{equation}
p^\mu\del_\mu f_q +   qF^{\alpha\beta}p_\beta\frac{\del f_q}{\del p^\alpha}=-\frac{p^\mu u_\mu}{\tau}\kl f_q-f_{\text{eq},q} \kr.\label{AndersonWittigModel}
\end{equation}
It allows for a straightforward calculation of the quark distribution $f_q$ after applying an external electric field. The gluon distribution remains thermal $f_g=f_{\text{eq},g}$ and is not affected by the electric field. In Appendix \ref{appsec:Relaxationsection} the electric conductivity is calculated from Eq.~\eqref{AndersonWittigModel} assuming very small electric fields, and no cross effects between heat and electric conductivity,
\begin{equation}
\sel=\frac{1}{3T}\sum_{k=1}^M q_k^2 n_k \tau.
\label{EasyRelaxationFormula}
\end{equation}
Here, $\tau$ is the mean time between collisions of particles, independent of the particle type. 
The Green-Kubo formula  Eq.~\eqref{WichtigeFormel}, integrated out, including the analytic variance Eq.~\eqref{eq:variance_analytic_main_text}, gives again exactly the same expression \eqref{EasyRelaxationFormula}. This justifies the identification of the inverse slope parameter $\tau$ of the exponential correlation function $C(t)$ as a mean time between collisions. It is essentially a parameter, describing the relaxation of the disturbed quark distribution towards the equilibrium solution. The smaller the value of $\tau$, the faster the disturbed system will relax back to equilibrium. It depends on the total (transport) cross section $\sigma_{\text{tot}}$ ($\sigma_{\text{tr}}$), and the total particle density $n_{\text{tot}}$. This can be phenomenologically parametrized as
\begin{equation}
\tau=\frac{1}{n_{\text{tot}}\sigma_{\text{tr}}}=\frac{3}{2n_{\text{tot}}\sigma_{\text{tot}}},
\label{eq:relaxation_times}
\end{equation}
thus
\begin{equation}
\sel=\frac{1}{2}\frac{\sum_{k=1}^M q_k^2 n_k}{\sum_{k=1}^M n_k}\frac{1}{T\sigma_{\text{tot}}}.\label{FinalEasyRelaxationFormula}
\end{equation}
Note that the relaxation time can be split up into different parts coming from the interactions amongst the different species ~\cite{Puglisi2014a}. For the electric conductivity, only $q-g$ and $q-\bar{q}$ scattering is relevant. 

The authors of Refs.~\cite{Cercignani,Kremer2003} use Eq.~\eqref{AndersonWittigModel} to derive an expression for the electric conductivity of an ultrarelativistic mixture of photons and electrons. Their expression for two components (electrons, subscript $e$; photons, subscript $\gamma$) reads
\begin{equation} 
\sel=\frac{q_e^2\tau_{e\gamma} n_e}{12nT}\left(3n_e+4n_\gamma\right).\label{Kremergleichung}
\end{equation}
This expression is calculated by taking only the partial heat flows of the electrons into account.
In principle we can translate Eq.~\eqref{Kremergleichung} into the case of several quarks and gluons. However, in the scenario shown in Fig.~\ref{BampsvsAnalytics}, the gluons scatter with the same cross section as the quarks, and $g-g$ collisions, in particular, can happen. In Ref.~\citep{Greif2013}, it was discussed that in this case gluons do have a finite heat flow. For this reason, we expect a deviation of Eq.~\eqref{Kremergleichung} (generalized to multiple particle species) with the numerical results from BAMPS.


In Fig.~\ref{BampsvsAnalytics}, we compare numerical results from BAMPS  with the relaxation time solution \eqref{FinalEasyRelaxationFormula} and the approach from Eq.~\eqref{Kremergleichung}. Note that the charges are explicitly multiplied out in the results using $e^2=4\pi/137$.
None of the formulas agrees perfectly with the numerical results. Equations \eqref{EasyRelaxationFormula} and \eqref{Kremergleichung} used strong approximations, simplifying the collision term. We expect a kinetic calculation using a linearized collision term, similar to Refs.~\cite{DeGroot,Denicol2012,Denicol2010}, to be closer to the numerical results. This analytic calculation has to our knowledge not been carried out yet.
In Refs.~\cite{Greif2013} and ~\cite{Wesp2011}, numerical results from BAMPS for the heat conductivity and shear viscosity were very close to analytic calculations from resummed transient relativistic fluid dynamics ~\cite{Denicol2012b} and derivations in the Navier-Stokes approximation ~\cite{DeGroot,Huovinen2009}.

We conclude by giving the numerically obtained precise Drude-type formula for the electric conductivity for an electrically neutral system of $k=\lbrace 1,\ldots,M\rbrace$ species of massless particles with electric charge $q_k$ and number density $n_k$ at temperature $T$, scattering with isotropic constant total cross section $\sigma_{\text{tot}}$:
\begin{equation}
\sel\left(\frac{1}{2}\frac{\sum_{k=1}^M q_k^2 n_k}{\sum_{k=1}^M n_k}\frac{1}{T\sigma_{\text{tot}}}\right)^{-1}=0.9\pm 0.01.
\end{equation}

\section{pQCD-based cross sections}
\label{sec:pqcd_systems_results}
\begin{figure}[t!]
  \centering
\includegraphics[width=\columnwidth]{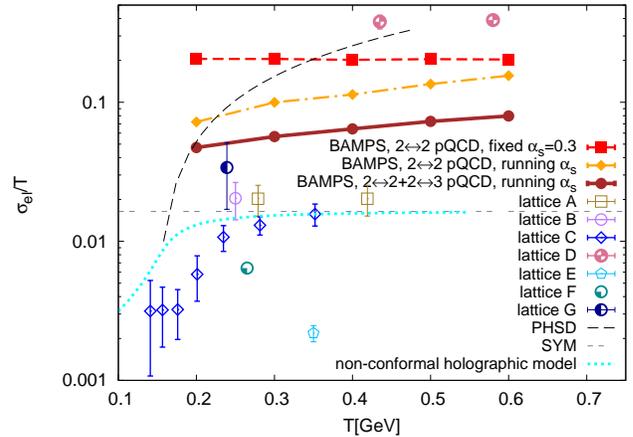}
\caption{
Numerical results for the electric conductivity (filled symbols) compared to recent results from literature. The open symbols represent results from lattice QCD. PHSD:  ~\cite{Cassing2013}, SYM: ~\cite{Starinets2006}, nonconformal holographic model: ~\cite{Finazzo2014}, lattice A: ~\cite{Aarts2007}, lattice B: ~\cite{Brandt2013}, lattice C: ~\cite{Amato2013a}, lattice D:  ~\cite{Gupta2004}, lattice E: ~\cite{Buividovich2010}, lattice F: ~\cite{Burnier2012}, lattice G: ~\cite{Ding2011}. The electric charge is explicitly multiplied out, $e^2=4\pi/137$. Around $T=0.3~\mathrm{GeV}$, results from Ref.~\cite{Qin2013} (not shown), using a Dyson-Schwinger approach, are consistent with the results from Ref.~\cite{Amato2013a}.  }\label{Superplot}
\end{figure}

The methods to extract the electric conductivity presented in Sec.~\ref{sec:NumericalMethods} are reliable, and can be readily applied to more realistic scenarios, where leading-order pQCD-based cross sections ~\cite{Uphoff2014a,Fochler2013,Xu2005} are employed. Figure~\ref{Superplot} depicts our results for the electric conductivity using pQCD cross sections.
The filled red squares are results for a scenario considering only elastic $2\leftrightarrow 2$ pQCD interactions when the strong coupling constant is fixed to $\alpha_s=0.3$. The ratio $\sel/T$ is constant within the small errors. This is expected by dimension, as leading-order pQCD cross sections $\sigma_{\text{pQCD}}(\alpha_s=0.3)$ behave typically as $\sim T^{-2}$, and $\sel/T\sim T^{-2}\sigma_{pQCD}^{-1}$. In Ref. ~\cite{Arnold2000}, the authors predict the same behavior with temperature. We emphasize that the results of Ref.~\cite{Arnold2000} are not directly comparable with our results, since we simulate a pure QCD plasma, and in Ref. ~\cite{Arnold2000} the electric current is assumed to be carried exclusively by leptons.
The filled yellow diamonds depict the electric conductivity over temperature for the same setup as before, but now the coupling constant $\alpha_s$ is running ~\cite{Uphoff2011a}. Evaluating the QCD running coupling at the momentum transfer of each microscopic interaction leads to an effective temperature dependence of the coupling ~\cite{Uphoff2014a}, and hence a qualitatively different temperature dependence of the electric conductivity is obtained. The interaction strength decreases with increasing temperature, and accordingly the effective cross section decreases.
The filled dark red circles are results for the most realistic scenario. Here we employ elastic $2\leftrightarrow 2$ and inelastic $2\leftrightarrow 3$ scatterings, and the running coupling $\alpha_s$. 
The LPM effect is modeled as described in Ref.~\cite{Uphoff2014a}, using the LPM parameter $X_{\text{LPM}}=0.3$. The result is sensitive to the LPM cutoff $X_\text{LPM}$, but its value is fixed by comparing BAMPS simulations of full heavy-ion collisions with experimental data for the nuclear modification factor; see Sec.~\ref{sec:BAMPS}. As an example, changing $X_{\text{LPM}}=0.3$ to $X_{\text{LPM}}=0.5$ or $X_{\text{LPM}}=1.0$ increases the electric conductivity by about $16\%$ or $40\%$. 
We emphasise again that the scattering rates of radiative processes are governed by the improved Gunion-Bertsch matrix elements, which were developed in Refs.~\cite{Fochler2013,Uphoff2014a}.
The inclusion of inelastic collisions accounts for an overall higher effective cross section than in the elastic scenarios. Therefore, the electric conductivity decreases by about $40\%$, and the slope of $\log(\sel/T)(T)$ decreases slightly.  
Nevertheless, the temperature dependence seems to be dominated by the running of $\alpha_s$.

This study allows us in a unique way to study the overall effective scattering rates for a hot QCD plasma microscopically, including all leading-order elastic and inelastic processes. The electric conductivity reflects in a profound way the effect of inelastic pQCD scattering and the running of $\alpha_s$. We believe that this is an important result of pQCD, and comparisons with other theories are reasonable.

In Fig.~\ref{Superplot}, we contrast the electric conductivity obtained using BAMPS with recent lQCD results, the transport model PHSD, a conformal, and a nonconformal holographic computation.
Comparison with lQCD data has to be taken with care. Obviously, published results from lQCD for the electric conductivity differ greatly, and general trends cannot be concluded, other than that most results lie within $0.001\leq\sel/T \leq 0.1$. The error bars are mostly large, or not quoted.
The presented results from the BAMPS transport simulation lie between $0.04\leq\sel/T\leq 0.08$ for temperatures $0.2~\mathrm{GeV}\leq T\leq 0.6~\mathrm{GeV}$.
The main differences amongst the lQCD setups are the QCD actions, different methods to handle the inversion problem and different numbers of dynamical and valence quarks.
It has to be mentioned, that the temperature, at which certain results are valid, is often quoted in units of the critical temperature. The precise value of the critical temperature requires, in turn, a lattice analysis. We omit at this point a further detailed comparison amongst the lQCD  results, which can be found elsewhere ~\cite{Meyer2011}.
The most recent results from lQCD are given by the authors of Ref.~\cite{Amato2013a}(open blue diamonds, dashed line to guide the eye). They provide the largest set of data for different temperatures so far, and use ensembles of 2+1 dynamical flavors. Their temperature dependence for $\sel/T$ above $T\sim 250~\mathrm{MeV}$ is similar to the results from BAMPS with running coupling. This qualitative agreement supports the physical validity of the implemented inelastic scattering processes of BAMPS. However, the results of Ref.~\cite{Amato2013a} are a factor $\sim 4$ smaller than ours.
In addition, we show in Fig.~\ref{Superplot} results from the PHSD transport approach by the black dashed line ~\cite{Cassing2013,Steinert2013}. One observes a significantly different temperature dependence. The value obtained in a conformal Super-Yang Mills plasma is shown by the constant grey dashed line ~\cite{Starinets2006}. The authors of  Ref.~\cite{Finazzo2014} used a nonconformal, bottom-up holographic model to compute the electric conductivity (cyan dotted line). Their model adequately describes recent lattice data for QCD thermodynamics at zero chemical potential.
\section{Conclusion and Outlook}
\label{conclusion}
In this work we extracted the electric conductivity coefficient for a dilute gas of massless and classical particles described by the relativistic Boltzmann equation. For this purpose we employed the microscopic transport model BAMPS in a static multipartonic system. We use two independent methods to extract the transport coefficient, and see nice agreement between the two. We present results using binary collisions and a constant isotropic cross section.  Here we find agreement with the relativistic generalization of the Drude formula for the electric conductivity in the functional dependence as well as the overall magnitude, with deviations of $\sim 10\%$. The Drude formula, being a relaxation time approximation is by no means expected to be exact. As further refined computations are lacking to date, we quote the new literature value for the electric conductivity for the class of systems described above.

Furthermore, we calculate the electric conductivity for systems in which the quarks and gluons scatter elastically and inelastically along leading-order pQCD cross sections. For fixed coupling and elastic pQCD collisions, we observe the expected constant temperature dependence of $\sel/T$. Running coupling is explicitly seen in a rise of $\sel/T$ with temperature. Inelastic collisions account for an overall decrease of the conductivity when compared to an elastic scenario. Finally, including all inelastic collisions and running coupling we obtain the electric conductivity of the QGP within the partonic transport simulation BAMPS.
Here we employ the recently developed improved Gunion-Bertsch matrix element, and the effective modelling of the LPM effect. 

Comparisons with lQCD computations is in general a difficult task; however, we see a very similar functional dependence of the electric conductivity on temperature compared to recent lQCD calculations. 
The electric conductivity opens up important possibilities to learn about the interaction properties of the QGP. In the future, more refined analytic and lQCD calculations, compared with microscopic transport simulations, can further restrain the value of the electric conductivity. To this end, it will shed light upon the microscopic interaction inside the QGP.
\begin{acknowledgments}
The authors are thankful to V. Greco and A. Puglisi for fruitful discussions. The authors thank G.S. Denicol, H. van Hees, J. Uphoff, and F.Senzel for constant interest in the subject and helpful discussions. The authors are grateful to the Center for Scientific Computing (CSC) Frankfurt for the computing resources. M.G. is grateful to the ``Helmhotz
Graduate School for Heavy Ion Research''. I.B. acknowledges support by BMBF. 
Z.X. is supported by the NSFC and the MOST under Grants No. 11275103, No. 11335005, and No. 2014CB845400.
This work was supported by the Helmholtz International Center for FAIR within
the framework of the LOEWE program launched by the State of Hesse.
\end{acknowledgments}
\appendix
\section{Variance for the electric current density}
\label{appsec:Variancesection}
For a $K$-tuple $a=\left\lbrace x^a_1,x^a_2,\ldots,x^a_K\right\rbrace$ of identically distributed variables, the variance is defined as
\begin{equation}
\mathrm{Var}\left\lbrace\mathrm{a}\right\rbrace=\sum\limits_{i=1}^K\kl x^a_i - \bar{a} \kr^2 P(x^a_i)
\label{eq:varinance-def}
\end{equation}
where $P(y)$ denotes the probability to find the value $y$, and $\bar{a}$ is the statistical average of the ensemble. From the correlation function Eq.~\eqref{eq:classical_correlation_function}, taking the total electric current in the $l$ direction \eqref{eq:current_to_be_correlated} for the $K$  available  time steps  as $K$-tuple, we identify
\begin{equation}
C(0)=\frac{1}{K}\sum\limits_{i=1}^K j^l(t_i)j^l(t_i)=\mathrm{Var}\left\lbrace j^l(t)\right\rbrace.
\end{equation}
As previously done in Refs.~\cite{Wesp2011,WespThesis} for the case of shear viscosity, it is possible to calculate the variance of the electric current density analytically. This is very useful when compared to numerical results.
Using the fact that for linear functions of the stochastic variable ($\alpha,\beta \in \mathbb{R}$), 
\begin{equation}
\mathrm{Var}\left\lbrace \alpha a+\beta \right\rbrace=\alpha^2\mathrm{Var}\left\lbrace a \right\rbrace,
\label{eq:lintrafo}
\end{equation}
and for $L$ uncorrelated $K$-tuples $\left\lbrace a_1,a_2,\ldots,a_L\right\rbrace$, 
\begin{equation}
\mathrm{Var}\left\lbrace\sum\limits_{i=1}^L a_i \right\rbrace = \sum\limits_{i=1}^L\mathrm{Var}\left\lbrace a_i \right\rbrace,
\label{eq:bien}
\end{equation}
we obtain with Eq.~\eqref{eq:current_to_be_correlated} for systems with $M$ particle species (and  $N_k$ particles of species $k$) the variance of the current in the $l$ direction,
\begin{align}
\mathrm{Var}\left\lbrace j^l(t) \right\rbrace&=\sum\limits_{k=1}^M \frac{q_k^2}{V^2}N_k\mathrm{Var}\left\lbrace \frac{p^l}{p^0} \right\rbrace\n
&=\sum\limits_{k=1}^M \frac{q_k^2}{V} \frac{n_k}{3},
\label{eq:variance_analytic}
\end{align} 
assuming isotropic fluctuations. 
%
\section{Relaxation-Time approximation for the electric conductivity}
\label{appsec:Relaxationsection}
In this appendix we use the Anderson-Witting model equation ~\cite{Anderson1974} to derive directly an expression for the electric conductivity. We assume for simplicity that there are as many quarks (charge $q$) as antiquarks (charge $-q$) of each flavor, and assume the presence of uncharged gluons. All particles are massless.
The equilibrium distribution function of quark species $k$ is
\begin{equation}
f_\text{eq,k}=g_k e^{-\beta p^0},
\end{equation}
where $g_k$ is the degeneracy. We  investigate the effect of an external, small and static electric field.
It will bring the quark distribution slightly off equilibrium, whereas the gluon distribution is exactly in equilibrium.
The Boltzmann equation in the relaxation-time approach of Anderson-Witting ~\cite{Anderson1974} reads
\begin{equation}
p^\mu\del_\mu f_k +   qF^{\alpha\beta}p_\beta\frac{\del f_k}{\del p^\alpha}=-\frac{p^\mu u_\mu}{\tau}\kl f_k-f_{\text{eq},k} \kr,\label{relaxation-time-formula}
\end{equation}
where $f_k=f_k(x,\vec{p},t)$ denotes the full distribution function of species $k$, and the mean time between collisions $\tau$ is given by Eq.~\eqref{eq:relaxation_times}.
We assume that the distribution function of the quarks is always close to equilibrium,
\begin{equation}
f_k(x,\vec{p},t)=f_\text{eq,k}+f_\text{eq,k}\phi_k.
\end{equation}
The field strength tensor $F^{\mu\nu}$ can be expressed through the electric field and the magnetic flux tensor, which is directly related to the magnetic induction, 
\begin{equation}
F^{\mu\nu}=u^\nu E^\mu - u^\mu E^\nu - B^{\mu\nu}.
\end{equation}
Our task is to investigate the influence of an electric field on the medium, so the magnetic induction is set to zero, $B^{\mu\nu}\equiv 0$. Note that $E^0=0$ and $E^i$ are the components of the electric field in the LRF of the fluid.
The electric current density of species $k$ in the $x$ direction is
\begin{equation}
j^x_k=q_k\int \frac{d^3\vec{p}}{(2\pi)^3}\, \frac{p^x}{p^0}f_k=g_k\tau \frac{8}{3} \frac{\pi q_k^2}{(2\pi)^3\beta^2}E^x.
\end{equation}
We can read off the electric conductivity $j^x=\sel E^x$ using the relaxation time Eq.~\eqref{eq:relaxation_times} and the particle density of species $k$, $n_k=g_k T^3/\pi^2$. Then we generalize to several species by the replacement $g_k q_k^2 \rightarrow \sum_k g_k q_k^2$:
\begin{equation}
\sel=\frac{8}{3}\frac{\sum_k g_k q_k^2}{8\pi^3}T^2\tau=\frac{1}{2}\frac{\sum_k g_k q_k^2}{\sum_k g_k}\frac{1}{T\sigma_{\text{tot}}}\label{FinalRelax}.
\end{equation}
We note, that this result can also be obtained from Eq.~\eqref{Non-relativistic-Drude} by the replacements $m_q \rightarrow 3T$ and $n_q q^2 \rightarrow \sum_k n_k q_k^2$, 
\begin{equation}
\sel=\frac{q^2\tau n_q}{m_q}\ \longrightarrow\ \sel=\frac{1}{2}\frac{\sum_k n_k q_k^2}{\sum_k n_k}\frac{1}{T\sigma_{\text{tot}}}.
\end{equation} 

\bibliographystyle{apsrev4-1}
\bibliography{library_manuell.bib}

\end{document}